\documentclass[11pt]{article}
\usepackage[english]{babel}
\usepackage[latin1]{inputenc}
\usepackage{latexsym,amsfonts,amsmath,amssymb,euscript,epsfig,dsfont,graphicx,color,float}
\usepackage[active]{srcltx}
\usepackage[T1]{fontenc}
\usepackage{url}

\title{Nash equilibria with partial monitoring;\\ Computation and Lemke-Howson algorithm.}
\author{Vianney Perchet \thanks{Laboratoire de Probabilit\'es et de Mod\`eles Al\'eatoires, Universit\'e Paris 7, 175 rue du Chevaleret, 75013 Paris. vianney.perchet@normalesup.org}}

\DeclareMathOperator{\argmax}{argmax}

 \DeclareMathOperator{\BR}{BR}
\DeclareMathOperator{\co}{co}

\newcommand\cX{\mathcal{X}}
\newcommand\cY{\mathcal{Y}}
\newcommand\cA{\mathcal{A}}
\newcommand\cK{\mathcal{K}}
\newcommand\cU{\mathcal{U}}
\newcommand\cL{\mathcal{L}}
\newcommand\cI{\mathcal{I}}
\newcommand\cH{\mathcal{H}}

\newcommand\cM{\mathcal{M}}
\newcommand\cG{\mathcal{G}}

\newcommand\cP{\mathcal{P}}
\newcommand\cB{\mathcal{B}}

\newcommand\R{\mathds{R}}

\newcommand\E{\mathds{E}}

\newcommand\bH{\mathbf{H}}
\newcommand\bM{\mathbf{M}}

\begin{document}
\maketitle
\newcounter{compteur}
\newtheorem{proposition}[compteur]{Proposition}
\newtheorem{theorem}[compteur]{Theorem}
\newtheorem{lemma}[compteur]{Lemma}
\newtheorem{corollary}[compteur]{Corollary}
\newtheorem{hypo}{Assumption}
\newtheorem{definition}{Definition}
\newtheorem{remark}{Remark}
\newtheorem{example}{Example}

\begin{abstract}
In two player bi-matrix games with partial monitoring,  actions played are not observed, only some messages are received. Those games satisfy a crucial property of usual bi-matrix games: there are only a finite number of required (mixed) best replies. This is very helpful while investigating sets of Nash equilibria: for instance, in some cases, it allows to relate it  to the set of equilibria of some auxiliary game with full monitoring.

In the general case, the Lemke-Howson algorithm is extended and, under some genericity assumption,  its output are Nash equilibria of the original game. As a by product, we obtain an oddness property on their number.
\end{abstract}

\section*{Introduction}
In finite games, proving the existence of Nash equilibria~\cite{Nas50,Nas51} is not very challenging, as they are fixed points of some correspondence. On the other hand, computing the whole set of Nash equilibria  (or exhibiting some of its topological properties) is quite hard~\cite{Pap07}. Similar statements can be made in games where actions chosen  or  actual payoff mappings are (partially) unknown. These games are getting increasing interest and have been referred as \textsl{robust}~\cite{AghBer06},  \textsl{ambiguous}~\cite{Bad10}, with \textsl{uncertainty}~\cite{Kli96}, \textsl{partially specified}~\cite{Leh07}, and so on. Indeed, Nash equilibria are defined similarly as fixed points of some complicated --~yet regular -- correspondence;  existence is then ensured, almost always using the very same argument of Nash~\cite{Nas50},  Kakutani's fixed point theorem. So the focus shall not be  existence, but characterizations and computation of these equilibria.

In full generality, and as expected as it is a more complex set-up, this turns out to be a very challenging problem~\cite[Section 5]{AghBer06}. We therefore consider here the class of bi-matrix games with \textsl{partial monitoring}, see e.g.,~\cite{MerSorZam94}, which contains all  two-player finite games. In this framework, players might not observe perfectly their opponent's actions (yet we always assume that one knows his own choice); they only receive messages. Depending on the game, actions and messages can in fact  be  correlated as well as independent; we could even assume that the latter is random, but up to some lifting, this can be reduced to the deterministic case, see~\cite{Per12}. These games are therefore described by two pair of matrices: a first pair for  payoffs  and a second pair for  messages received.

Players, facing uncertainties upon their payoffs, cannot directly maximize them. As it is usual now~\cite{GilSch89,BenElgNem09}, we assume that they  optimize their behavior with respect to  the  worst possible scenario, leading to \textsl{maxmin expected utility}.

\bigskip

Using topological properties of linear mappings and projection, we recover surprisingly the following  fundamental property of finite bi-matrix games with full monitoring (when actions are observed). There exists a fixed finite subset of (mixed) actions  containing  best-replies to any action of the opponent. While obvious with full monitoring by considering whole set of pure actions, this result is not immediate with partial monitoring (and actually  incorrect in another class of games than the one considered here).

In the subclass of games called with \textsl{semi-standard} information structure,  developed in Section~\ref{SE:SemiStand}, this allows the construction of  an auxiliary game with full monitoring such that its Nash equilibria are (in some sense) also equilibria of the original game. So any property  with full monitoring holds for this type of games.

In the general case, this direct reduction is incorrect. Yet we prove  in Section~\ref{SectionLHGeneral} that Nash equilibria satisfy again another usual properties of full monitoring, see~\cite{Sha74}. Using this, sets of Nash equilibria are characterized and some of them can be computed  using the Lemke-Howson algorithm~\cite{LemHow64},  recalled briefly in Section~\ref{SectionLHalgo}. These computations are illustrated in Section~\ref{examplesemistandardnecsuite}; other claims are also, as often as possible, accompanied  by examples. Interestingly, since Nash equilibria -- even with partial monitoring -- are end-points of a special instance of the Lemke-Howson algorithm, some oddness property of their set is preserved (as soon as some genericity assumption is satisfied).

\section{Two players game with partial monitoring}

Consider a finite two players game $\Gamma$ where action of  player 1 (resp.\ player 2) is  by $\cA$ (resp.\ $\cB$) and his payoff mapping is $u: \cA \times \cB \to \R$ (resp.\ $v: \cA \times \cB \to \R$), extended multi-linearly to $\cX\times\cY$. We denote by $\cX=\Delta(\cA)$ and $\cY=\Delta(\cB)$ mixed action sets of both players. We also assume that they have partial monitoring: they do not observe  actions of their opponent but  receive messages instead, see~\cite{MerSorZam94}. Formally, there exist two convex compact  sets of messages $\cH$ and $\cM$ and two signaling mappings  $H$ and $M$  from $\cA\times\cB$ into $\cH$ or $\cM$ (also extended multi-linearly) such that if players choose $x \in \cX$ and $y \in \cY$,  player 1 gets a  payoff of $u(x,y)$ but he only observes the message $H(x,y) \in \cH$. On his side, player 2 gets a payoff of $v(x,y)$ and he observes $M(x,y) \in \cM$. 

\medskip

No matter his choice of actions, player 1 cannot distinguish between $y$ and $y' \in \cY$  satisfying  $H\left(a,y\right)=H\left(a,y'\right)$ for every $a \in \cA$.  We thus define the \textsl{maximal informative mapping} $\bH : \cY  \to \cH^{A}$ ($A$ stands for the cardinality of $\cA$) by:
\[\forall\, y \in \cY, \ \bH(y)=\Big[H\left(a,y\right)\Big]_{a \in \cA} \in  \cH^{A}.\]
Similarly, the maximal informative payoff of  player 2, $\bM: \cX  \to \cM^{B}$, is defined by \[\forall\, x \in \cX, \ \bM(x)=\Big[M\left(x,b\right)\Big]_{b \in \cB} \in  \cM^{B}.\]
These linear mappings induce  \textsl{uncertainty correspondences} $\Phi:  \cY \rightrightarrows \mathds{R}^{A}$ and $\Psi:  \cX \rightrightarrows \mathds{R}^{B}$ defined by:
\[\Phi\left(y\right)= \left\{u(\cdot,y') \in \mathds{R}^{A};\ \bH\left(y'\right)=\bH\left(y\right)\right\} \ \text{and}\ \Psi\left(x\right)= \left\{v(x',\cdot) \in \mathds{R}^{B};\ \bM\left(x'\right)=\bM\left(x\right)\right\}.\]
Informally, if player 2 chooses $y \in \cY$, then player 1 cannot distinguish it from any other $y'$ that have the same image under $\bH$; thus, if he plays  $x \in \cX$, he cannot compute his actual payoff as he only infer that it will be on the form $\langle x , U \rangle$ for some unknown $U$ that must belong to $\Phi(y)$ ( which is also equal to $\Phi(y')$).

When dealing with uncertainties, best replies  are extended, following~\cite{GilSch89,AghBer06}, into
\[\BR_1: \cP(\R^{A}) \rightrightarrows \cX \ \ \text{with}\ \ \BR_1(\mathcal{U})=\arg\max_{x \in \cX} \inf_{U \in \mathcal{U}} \langle x,U\rangle\ ,\]
where $ \cP(\R^{A})$ is the family of subsets of  $\R^{A}$. This is well-defined since  $x \mapsto \inf_{U \in \mathcal{U}} \langle x,U\rangle$ is concave and upper semi-continuous hence  maxima are attained. $\BR_2:\cP(\R^{B}) \rightrightarrows \cY$ is defined in a similar way. Definition~\ref{DF:NE} below of Nash equilibria with partial monitoring (see also~\cite{Per12} for more details and explanations) follows naturally.

\begin{definition}\label{DF:NE}
 $(x^*,y^*) \in  \cX\times \cY$ is a Nash equilibrium of a game with partial monitoring iff $x^* \in \BR_1(\Phi(y^*))$ and $y^*\in \BR_2(\Psi(x^*))$, i.e., iff
 \[ x^* \in \arg\max_{x \in \cX} \inf_{U \in \Phi(y^*)} \langle x,U\rangle\ \ \text{and} \ y^* \in \arg\max_{y \in \cY} \inf_{V \in \Psi(x^*)} \langle x,V\rangle\ .
 \]  
 \end{definition}

\section{A warm-up: semi-standard structure}\label{SE:SemiStand}

We first consider an easy case: games with a \textsl{semi-standard} information structure. Informally, it implies that  action sets  are partitioned into subsets of undistinguishable  actions (but it is always possible to distinguish between these subsets).
\begin{definition}\label{DF:SemiStandard}
The information of  player 1 (and similarly for player 2) is semi-standard if there exists a partition $\{\cB_i;\ i\in \cI\}$ of $\cB$ such that
\begin{itemize}
\item[i)]{If $b$ and $b'$ belong to the same cell $\cB_i$ then  $\bH(b)=\bH(b')=\bH_i$ and}
\item[ii)]{The family $\{\bH_i;\ i \in \cI\}$ is linearly independent, i.e. if $\sum_{i \in \cI} \lambda_i \bH_i=\sum_{i \in \cI} \gamma_i \bH_i$ then $\lambda_i$ and $\gamma_i$ must be equal, for every $i \in \cI$.}
\end{itemize}
A game has a semi-standard structure if both $\bH$ and $\bM$ satisfy these properties. \end{definition}

In particular, this means that, for every $y \in \cY$, given $\bH(y) \in \cH^{A}$, player 1 can only infer $\{y_i; i \in \cI \}$ where $y_i=\sum_{b \in \cB_i} y[b]$ is the probability (accordingly to $y$) of choosing an action in $\cB_i$.

\begin{example}
If $\cH=[0,1]^d$ and, no matter $b\in \cB$,  $H(a,b)=H(a',b)=e_b$ where $e_b$ is a vector with only one non-zero coordinate which is 1, then player 1 has a semi-standard information structure. However, if we do not assume that $H(a,b)=H(a',b)$, then this is no longer true. 

Indeed, let $\cA=\{a,a'\}$, $\cB=\{b_1, b_2,b_3,b_4\}$, $\cH=[0,1]^2$ and $H$ be represented as
\begin{center}
\begin{tabular}{p{0.2cm}cp{1cm}p{7cm}}
$H$:&
\begin{tabular}{c|c|c|c|c|}
\multicolumn{1}{c}{}&\multicolumn{1}{c}{$b_1$}&\multicolumn{1}{c}{$b_2$}&\multicolumn{1}{c}{$b_3$}&\multicolumn{1}{c}{$b_4$}\\
\cline{2-5}
$a$ & $e_1$ & $e_2$ &$e_1$ &$e_2$ \\
\cline{2-5}
$a'$ & $e_1$ & $e_2$ &$e_2$ &$e_1$ \\
\cline{2-5}\end{tabular}
&& with $e_1=(1,0)$ and $e_2=(0,1)$.
\end{tabular}
\end{center}
The decomposition of point i) of Definition~\ref{DF:SemiStandard} must be $\bH_1=(e_1,e_1)$, $\bH_2=(e_2,e_2)$ and so on. However,  point ii) of the same definition is not satisfied since \[\bH\left(\frac{b_1+b_2}{2}\right)=\left(\frac{e_1+e_2}{2},\frac{e_1+e_2}{2}\right)=\bH\left(\frac{b_3+b_4}{2}\right).\]
\end{example}

In this framework,  following Lemma~\ref{LM:LinearSemiStandard} allows an easy reduction  from partial to full monitoring. But we need to recall first  the general concept of \textsl{polytopial complex} (a \textsl{polytope} is the convex hull of a finite number of points\footnote{A polytope can also be defined, in a totally equivalent way, as a compact and non-empty intersection of a finite number of half-planes}) on which our results rely:
\begin{definition}
A finite set $\{P_k;\ k \in \cK \}$  is a polytopial complex of a polytope $P \subset \mathds{R}^d$ with non-empty interior if:
\begin{itemize}
\item[i)]{For every $k \in \cK$, $P_k \subset P$ is a polytope with non empty interior;}
\item[ii)]{The union $\bigcup_{k \in \cK} P_k$ is equal to $P$;}
\item[iii)]{Every intersection of two differents polytopes $P_k\cap P_{k'}$ has an empty interior.}
\end{itemize}
\end{definition}

The following Lemma~\ref{LM:LinearSemiStandard} is an adaptation of an argument stated in~\cite[Theorem 34]{Per11}.
\begin{lemma}\label{LM:LinearSemiStandard}
There exists a finite subset $\{x_\ell;\ \ell \in \cL \}$ of $\cX$ that contains, for every $y \in \cY$, a maximizer of the program $\max_{x \in \cX} \min_{U \in \Phi(y)} \langle x, U \rangle$ and such that its convex hull contains the whole set of maximizers. Moreover, there exists a polytopial complex $\{\cY_\ell;\ \ell \in \cL\}$ of $\cY$ such that, for every $\ell \in \cL$,  $x_\ell$ is a maximizer on $\cY_\ell$.

Similarly, we denote by $\{y_k;\ k \in \cK\}$ the set defined in a dual way for player 2. \end{lemma}
\textbf{Proof:} Define, for every $i \in \cI$, the set of compatible outcomes with $\bH_i$ by:
\[\mathcal{U}_i=\left\{u(\cdot,y)\in \R^A;\ y \mbox{ s.t. } \bH(y)=\bH_i \right\}=\co \left\{u(\cdot,b);\ b \in \cB_i \right\}\ ,\]
where $\co$ stands for the convex hull; in particular, $\cU_i=\Phi(b)$, for all $b \in \cB_i$ and it is a polytope. So the mapping $\Phi$ is linear on $\cY$ since\footnote{Actually, the semi-standard structure could also be defined through the linearity of $\Phi$.} it is defined, for every $y \in \cY$, by
\[\Phi(y)=\sum_{i \in \cI} y_i \mathcal{U}_i = \sum_{i \in \cI} \sum_{b \in \cB_i} y[b] \mathcal{U}_i=\sum_{b\in \cB} y[b] \Phi(b).\]
Given $x^* \in \cX$ and $y \in \cY$, if $U^\sharp \in \Phi(y)$ is a minimizer  of $\min_{U \in \Phi(y)} \langle x^*, U\rangle$ then it can be assumed that $U^\sharp$ is a vertex of $\Phi(y)$, because a linear program is always minimized on a vertex of the admissible polytope. And necessarily $-x^*$ must belong to the normal cone to $\Phi(y)$ at $U^\sharp$~\cite[Theorem 27.4, page 270]{Roc70}. As a consequence, $-x^*$ must belong to the intersection of  $-\cX$ and a normal cone; more precisely, since $\langle x, U^\sharp \rangle$ is linear, $-x^*$ must be one of the vertices (or a convex combination of them) of this intersection.

However,  $\Phi(\cdot)$ is linear on $\cY$, so  normal cones at vertices -- their set is  called  normal fan -- are constant, see~\cite[Example 7.3, page 193]{Zie95} and~\cite[page 530]{BilStu92}. As a consequence, there exists a finite number of intersection between $-\cX$ and normal cones and they all have a finite number of vertices. The set of every possible vertices is denoted by $-\{x_\ell;\ l \in \cL\}$ and it always contains a maximizer (and any maximizer must belong to its convex hull).

\medskip

Since $\Phi$ is linear,  $y \mapsto \min_{U \in \Phi(y)} \langle x_\ell, U \rangle $ is also linear, for every $\ell \in \cL$; so $x_\ell$ is a maximizer on a polytopial subset of $\cY$. $\hfill \Box$

\begin{remark}
Lemma~\ref{LM:LinearSemiStandard} might be surprising to reader familiar with linear programming. Indeed, it is quite clear that if $u_1(\cdot,y)$ is linear then it is always maximized at one of the vertices of $\cX$. However, in our case,  $\min_{U \in \Phi(y)} \langle \cdot , U \rangle$ is not linear but only concave. So  it can be maximized anywhere in $\cX$, even in its interior.

So without some regularity of $\Phi$, the result would obviously e wrong. The key point of the proof is that, in our  framework, $\Phi$ is itself induced by the minimization of another linear mapping.  Lemma~\ref{LM:LinearSemiStandard} holds because  $\min_{U \in \Phi(y)} \langle \cdot , U \rangle$ is not just any concave mapping, but it has this extra specific property.\end{remark}

We now introduce an auxiliary game $\widetilde{\Gamma}$, with full monitoring, such that its Nash equilibria somehow coincide with Nash equilibria of $\Gamma$, the original game. Respective action sets of player 1 and 2 are $\cL$  and $\cK$ and   payoff mappings 
\[ \widetilde{u}(\ell,k)=\min_{U \in \Phi(y_k)} \langle x_\ell, U \rangle \quad \mbox{ and } \quad \widetilde{v}(\ell,k)=\min_{V \in \Psi(x_\ell)} \langle y_k, V \rangle.\] Any pair of mixed actions $(\mathbf{x},\mathbf{y}) \in \Delta(\cL)\times\Delta(\cK)$  induces a pair  $(x,y) \in \cX\times\cY$  defined by $x = \E_{\mathbf{x}}[x_\ell] \in \cX$. This means that, for every $a \in \cA$, the weight put by $x$ on $a$ is $x[a]:=\sum_{\ell \in \cL} \mathbf{x}[\ell]x_\ell[a]$; similarly, $y$ is defined by $y = \E_{\mathbf{y}}[y_k] \in \cY$.

\begin{theorem}\label{theosemi}
Every  Nash equilibrium of $\widetilde{\Gamma}$ induces a Nash equilibrium of $\Gamma$ and, reciprocally, every  Nash equilibrium of  $\Gamma$ is induced by a Nash equilibrium of  $\widetilde{\Gamma}$.
\end{theorem}
\textbf{Proof:} Let $(\mathbf{x}, \mathbf{y})$ be a Nash equilibrium of $\widetilde{\Gamma}$ and $(x,y)$ the induced mixed actions. By linearity of $\Phi$, one has $\sum_{k \in \cK} \mathbf{y}[k] \Phi(y_k) = \Phi(y)$ thus
\begin{align*} \widetilde{u}(\mathbf{x},\mathbf{y})&= \sum_{\ell \in \cL} \mathbf{x}[\ell] \sum_{k \in \cK} \mathbf{y}[k] \widetilde{u}(\ell,k)= \sum_{\ell \in \cL}\mathbf{x}[\ell] \sum_{k \in \cK}\mathbf{y}[k] \min_{U_k \in \Phi(y_k)} \langle x_\ell, U_k \rangle \\
&= \sum_{\ell \in \cL}\mathbf{x}[\ell] \min_{U \in \sum_{k \in \cK}\mathbf{y}[k] \Phi(y_k)} \langle x_\ell,U\rangle=    \sum_{\ell \in \cL}\mathbf{x}[\ell] \min_{U \in \Phi(y)} \langle x_\ell, U \rangle\\
&\leq \min_{U \in \Phi(y)} \left\langle \sum_{\ell \in \cL} \mathbf{x}[\ell] x_\ell, U \right\rangle= \min_{U \in \Phi(y)} \langle x, U \rangle.\end{align*}
Therefore with, respectively, the fact that $(\mathbf{x},\mathbf{y})$ is a  Nash equilibrium, the linearity of~$\Phi_1$ and Lemma~\ref{LM:LinearSemiStandard}, this implies that
\[\min_{U \in \Phi(y)} \langle x, U \rangle \geq   \widetilde{u}(\mathbf{x},\mathbf{y})\geq \max_{\ell \in \cL} \widetilde{u}(\ell,\mathbf{y})=\max_{\ell \in \cL}\min_{U \in \Phi(y)} \langle x_\ell, U \rangle=\max_{x' \in \cX}\min_{U \in \Phi(\mathbf{y})} \langle x', U \rangle.\]
Hence we have proved that $x \in \BR_1(\Phi(y))$; similarly $y \in \BR_2(\Psi(x))$, so $(x,y)$ is a Nash equilibrium of $\Gamma$.

\bigskip

Reciprocally, let $(x,y)$ be a Nash equilibrium of $\Gamma$. Lemma~\ref{LM:LinearSemiStandard} implies that $x$ is a convex combinations  of mixed actions in $\{x_\ell; l \in L\}$ that maximize $\min_{U \in \Phi(y)}  \langle x_\ell,U \rangle$. Denote by  $\mathbf{x} \in \Delta(\cL)$ this convex combination and  define $\mathbf{y}$ in a dual way.

Since $\mathbf{y} \in \Delta(\cK)$ induces  $y$, then one has, for every $\ell' \in \cL$:
\[ \widetilde{u}(\ell',\mathbf{y}) \leq \max_{x' \in \cX} \min_{U \in \Phi(y)} \langle x', U \rangle=\sum_{\ell \in \cL} \mathbf{x}[\ell] \min_{U \in \Phi(y)} \langle x_\ell, U \rangle =  \widetilde{u}(\mathbf{x},\mathbf{y})\ ,\]
where we used respectively the linearity of $\Phi$, the fact that $\mathbf{x}_\ell>0$ if $x_\ell$ is a maximizer and again the linearity of $\Phi$. Therefore $\mathbf{x}$ is a best reply   to $\mathbf{y}$  and the converse is true by symmetry: $(\mathbf{x},\mathbf{y})$ is a Nash equilibrium of $\widetilde{\Gamma}$. $\hfill \Box$

\medskip

Theorem~\ref{theosemi} implies that one just has to compute the set of Nash Equilibria of $\widetilde{\Gamma}$ in order to describe the set of Nash equilibria of $\Gamma$. For example, one might consider the Lemke-Howson algorithm ~\cite{LemHow64}  -- or LH-algorithm for short -- recalled briefly in the following section. 

If  $\widetilde{\Gamma}$  satisfies some \textsl{non-degeneracy} assumption,  the LH-algorithm  outputs a subset of Nash equilibria of both $\widetilde{\Gamma}$ and $\Gamma$. The specific assumption and how to  modify and apply this algorithm to any game are detailed in~\cite{von02}. 

\section{Quick reminder on Lemke-Howson algorithm}\label{SectionLHalgo}

The Lemke-Howson algorithm of~\cite{LemHow64} is designed to compute Nash equilibria of a two-player finite game with full monitoring. It is based on the decomposition of $\cX$ and $\cY$ into best-replies areas. Recall that $\cY_a:=\Big\{y \in \cY \ \text{s.t.}\ a \in \argmax_{a' \in \cA} u(a',y)\Big\} \subset \cY$, for any $a \in \cA$,  is the $a$-th best-reply area of player 1. The \textsl{genericity} assumption required by the LH-algorithm is the following:
\begin{hypo}\label{hypoLH}
$\{\cY_a;\ a \in \cA\}$  forms a polytopial complex of $\cY$ and  any $y \in \cY$ belongs to at most $m_y$ best reply areas $\cY_a$, where  $m_y$ is the size of the support of $y$. The similar condition  holds for $\{\cX_B;\ b \in \cB \}$.
\end{hypo}
Stated otherwise, Assumption~\ref{hypoLH} means that every $y \in \cY$ has at most $m_y$ best replies.

\bigskip

Each $\cY_a$ is a polytope, so denote by $V_2$ and $E_2$ the  set of all vertices and edges of these sets (necessarily $\cB \subset V_2$).  For technical purpose,  we also assume that $V_2$ contains another (abstract) point $0_2$  such that $(0_2,b)$ belongs to  $E_2$ for every $ b \in \cB$. This defines a graph $\cG_2=(V_2,E_2)$ over $\cY$ and similarly a graph $\cG_1=(V_1,E_1)$ over $\cX$. To each vertex $v_2\in V_2$ (and to each $v_1 \in \cB_1$) is associated the following set of labels:
 \[ L(v_2):=  \Big\{a \in \cA\ \mbox{s.t.}\ v_2 \in \cY_a\Big\} \bigcup \Big\{b \in \cB\ \mbox{s.t.} \ v_2[b]=0\Big\}  \subset \cA \cup \cB,\]
i.e., its best replies and pure actions on which it does not put any weight.  Label sets of abstract points $0_1$ and $0_2$ are $L(0_1)=\cA$ and $L(0_2)=\cB$.

This induces a product labelled graph $\cG_0=(V_0,E_0)$ over $\cX \times \cY$, whose set of vertices  is the cartesian product $V_0=V_1\times V_2$ and such that there exists an edge  in $E_0$ between $(v_1,v_2)$ and $(v'_1,v'_2)$  if and only if $v_1=v'_1$ and $(v_2,v'_2) \in E_2$ or $v_2=v'_2$ and $(v_1,v'_1) \in E_1$. The set of labels of $(v_1,v_2)$ is $L(v_1,v_2)=L(v_1) \cup L(v_2)$.

\medskip

 Nash equilibria are exactly \textsl{fully labeled} pairs $(v_1,v_2)$, i.e., if $L(v_1,v_2)=\cA_1 \cup \cA_2$; indeed, this means that an action $a$ is either not played (if $v_1[a]=0$) or  a best reply to $v_{2}$ (if $v_{2} \in \cY_a$). The LH-algorithm walks along edges of $\cG_0$, from vertices to vertices, and stops at a one of those points. We describe quickly in the remaining of this section how it works generically (i.e.\ for almost all games); for more details we refer  to~\cite{Sha74,von07}  and references therein.

\medskip

Starting at $v_0=(0_1,0_2)$ (which is fully labeled),  one label $\ell$ in $\cA \cup \cB$ is chosen arbitrarily. The LH algorithm visits sequentially \textsl{almost fully labeled} vertices $(v_t)_{t \in \mathds{N}}$ of  $\cG_0$, i.e., points such that $L(v_t) \supset \cA \cup \cB \backslash \{\ell\}$ and $(v_t,v_{t+1})$ is an edge in $E_0$. Generically, at any $v_t$ there exists at most one point  (apart from $v_{t-1}$) satisfying both properties, and  any end point must be fully labeled. 

As a consequence, when starting from any almost fully labeled point vertex, LH algorithm follows either a cycle (and stops when returning to a previously visited point) or a path whose endpoints are necessarily Nash equilibria (or $(0_1,0_2)$). This property can be used, for example, to prove that the number of Nash equilibria is generically odd.

\section{Characterization and computation of Nash equilibria}\label{SectionLHGeneral}
Without the semi-standard structure, Lemma~\ref{LM:LinearSemiStandard} and Theorem~\ref{theosemi} might not hold  since $\Phi$ is not linear (this is illustrated in  Example~\ref{EX:SansSemiStand}). However, we will show that, in the general case, we still have a similar property: $\Phi$ is piece-wise linear. This means that $\Phi$ is linear on a polytopial complex of $\cY$ (see the following Lemma~\ref{LM:LemmaPhiLinear}). Using this, it will be  easy to show (in Lemma~\ref{lemmalineargeneral} below) that best-replies areas forms a polytopial complex, allowing the generalization of LH-algorithm. Such decompositions have been recently used in related frameworks, see e.g.~\cite{vonZam10}.

\begin{example}\label{EX:SansSemiStand}
Assume that $\cA= \{T;B\}$, $\cB=\{L,C,R\}$ and $\cH=[0,1]$. Payoffs and player~1's message matrices  (player 2 has full monitoring) are given respectively  by:
\begin{center}
\begin{tabular}{p{0.2cm}cp{1cm}p{0.2cm}c}
u: & 
\begin{tabular}{c|c|c|c|}
\multicolumn{1}{c}{} & \multicolumn{1}{c}{$L$}&\multicolumn{1}{c}{$C$} & \multicolumn{1}{c}{$R$}\\
\cline{2-4}
$T$& $(1,1)$ & $(0,0)$ & $(0,0)$  \\
\cline{2-4}
$B$& $(0,0)$ & $(1,2)$& $(1,0)$ \\
\cline{2-4}
\end{tabular}
&& H: &
\begin{tabular}{c|c|c|c|}
\multicolumn{1}{c}{} & \multicolumn{1}{c}{$L$}&\multicolumn{1}{c}{$C$} & \multicolumn{1}{c}{$R$}\\
\cline{2-4}
$T$& $0$ & $1$ & $1/3$  \\
\cline{2-4}
$B$& $0$ & $1$& $1/3$ \\
\cline{2-4}
\end{tabular}
\end{tabular}
\end{center}
Player 1 cannot distinguish between the mixed action $2/3L+1/3C$ and the pure action~$R$.

Following  notations of Lemma~\ref{LM:LinearSemiStandard}, one has $\{x_\ell;\ \ell \in \cL \}=\{T, B, M\}$ where $M=1/2T+1/2B$ and\footnote{To be extremely rigorous, the pure action $R$ should be removed since it is never a best response.} $\{y_k;\  k \in \cK\}=\{L,C,R\}$. Thus $\widetilde{\Gamma}$ is defined by the following matrix:
\begin{center}
\begin{tabular}{c|c|c|c|}
\multicolumn{1}{c}{} & \multicolumn{1}{c}{$L$}&\multicolumn{1}{c}{$C$} & \multicolumn{1}{c}{$R$}\\
\cline{2-4}
$T$& $(1,1)$ & $(0,0)$&$ (0,0)$ \\
\cline{2-4}
$B$& $(0,0)$ & $(1,2)$ & $(1/3,0)$ \\
\cline{2-4}
$M$& $(1/2,1/2)$ & $(1/2,1)$ & $(1/2,0)$\\
\cline{2-4}
\end{tabular}
\end{center}
This game has three Nash Equilibria: $(T,L)$, $(B,C)$ and $(2/3 T + 1/3 B, 1/2 L + 1/2 C)$. Although the first two are indeed Nash equilibria of $\Gamma$, this is not true for the last one. Indeed, $\Phi(1/2L+1/2C)=\{(\lambda/2;1-\lambda_2);\ \lambda \in [0,1]\}$ and its best response is $\{T\}$.

Actually, and as we shall see in Example~\ref{examplesemistandardnecsuite},  $\Gamma$ has three Nash equilibria which are $(T,L)$, $(B,C)$ and $(1/3T+2/3M,3/4L+1/4C)=(2/3T+1/3B,3/4L+1/4C)$

\end{example}

\begin{lemma}\label{LM:LemmaPhiLinear}
The correspondence $\Phi$ is piecewise linear on $\cY$.
\end{lemma}
\textbf{Proof:} Since $\bH$ is linear from $\cY$ into $\cH^{A}$, then $\mu \mapsto \bH^{-1}(\mu)$ is piecewise linear on $\cH^{A}$, see~\cite[page 530]{BilStu92} and~\cite[Proposition 2.4, page 221]{RamZie96}.  Therefore,  by composition, $y \mapsto \bH^{-1}\big(\bH(y)\Big)$ is piecewise linear on $\cY$ and $y \mapsto u\bigg(\cdot,\bH^{-1}\big(\bH(y)\Big)\bigg)$ -- which is  by definition $\Phi$ -- is also piecewise linear on $\cY$.
$\hfill \Box$

\bigskip
So  Lemma~\ref{LM:LinearSemiStandard}  can be rephrased as follows.
\begin{lemma}\label{lemmalineargeneral}
There exists a finite subset $\{x_\ell;\ l \in \cL \}$ of $\cX$ that contains, for every $y \in \cY$, a maximizer of the program $\max_{x \in \cX} \min_{U \in \Phi(y)} \langle x, U \rangle$ and such that its convex hull contains the set of  maximizers. 

 Moreover, for every $\ell \in \cL$, $x_\ell$ is a maximizer on $\cY_l$ which is a finite union of polytopes.  Similarly, we denote by $\{y_k;\ k \in \cK\}$ and $\{\cX_k;\ k \in \cK\}$ the finite sets for player 2.
\end{lemma}
\textbf{Proof:} One just has to consider the polytopial complex $\{P_i;\ i \in \cI\}$ with respect to which $\Phi$ and $\Psi$ are piecewise linear and apply Lemma~\ref{LM:LinearSemiStandard} on each $P_i$. $\hfill \Box$

\medskip

Our main result is the following characterization of Nash equilibria in a general game with partial monitoring. We recall   that  $\mathbf{x} \in \Delta(\cL)$ induces the mixed action $x=\E_{\mathbf{x}}[x_\ell]$ where $x[a]$, the weight put by $x$ on $a \in \cA$, is equal to $\sum_{\ell \in \cL} \mathbf{x}[\ell] x_\ell[a]$.

\begin{theorem}\label{mainresult} Nash equilibria of $G_H$ are induced by points in $\Delta(\cL) \times \Delta(\cK)$ that are fully labelled  with respect to the two decompositions $\{\mathbf{Y}_\ell;\ \ell \in \cL\}$ and $\{\mathbf{X}_k;\ k \in \cK\}$  (and to the label set $\cL \cup \cK$)  defined by
\[\mathbf{Y}_\ell=\Big\{\mathbf{y} \in \Delta(\cK)\ \mbox{s.t.}\ \mathds{E}_{\mathbf{y}}[y_k] \in Y_\ell \Big\}=\left\{\mathbf{y} \in \Delta(\cK)\ \mbox{s.t.}\ x_\ell \in \arg\max_{\ell' \in \cL} \inf_{U \in \Phi(\mathds{E}_{\mathbf{y}}[y_k])}\  \langle x_{\ell'}, U\rangle \right\}\]
and similarly for $\mathbf{X}_k$.
\end{theorem}
\textbf{Proof:}
Consider any fully labelled point $(\mathbf{x},\mathbf{y}) \in \Delta(\cL) \times \Delta(\cK)$ and the induced mixed actions $x \in \cX$ and $y \in \cY$. By definition (see Section~\ref{SectionLHalgo}), for every $\ell \in \cL$ and $k \in \cK$, either $\mathbf{x}[\ell]=0$ or $\mathbf{y}$ belongs to $\mathbf{Y}_\ell$ (and similarly either $\mathbf{y}[k]=0$ or $\mathbf{x} \in \mathbf{X}_k$).

As a consequence, $x$ is a best reply to $y$ (and reciprocally) since:
\[ \min_{U \in \Phi(y)}\langle x,U\rangle=\min_{U \in \Phi(y)} \sum_{\ell \in \cL} \mathbf{x}[\ell] \langle x_\ell,U\rangle \geq \sum_{\ell \in \cL} \mathbf{x}[\ell] \min_{U \in \Phi(y)}  \langle x_\ell,U\rangle \geq \max_{x' \in \cX}\min_{U \in \Phi(y)}\langle x',U\rangle. \]

Therefore, any fully labelled point induces a Nash equilibrium of $\Gamma$.

\medskip

Reciprocally, if $(x,y)$  is a Nash equilibrium of $\Gamma$ then Lemma~\ref{lemmalineargeneral} implies that $x$ and $y$  belong to the convex hull of $\{x_\ell;\ \ell \in \cL\}$  and $\{y_k;\ k \in \cK\}$. More precisely, $x$ is a convex combination of the maximizers of $\min_{U \in \Phi(y)}\langle x_\ell, U \rangle$ (i.e.\ those $x_\ell$ such that $y \in Y_\ell$). If we denote this convex combination as $x=\sum_{\ell \in \cL} \mathbf{x}[\ell] x_\ell$, then necessarily either $\mathbf{x}[\ell]=0$ or $y$ belongs to $Y_\ell$ (and $\mathbf{y} \in \mathbf{Y}_\ell$). Therefore $(\mathbf{x},\mathbf{y})$ is fully labeled.
$\hfill \Box$

\bigskip

It remains to describe why the LH algorithm can be used in this framework. First, recall that every set  $Y_\ell$ or $X_k$ provided by Lemma~\ref{lemmalineargeneral} is a finite union of polytopes. So, up to an arbitrary subdivision of these non-convex unions (associated with maybe a duplication of some mixed actions, see  Example~\ref{examplefinal} below), we can assume that $\{Y_\ell;\ \ell \in \cL\}$ and $\{X_k;\ k \in \cK\}$ are finite families of polytopes. 

\begin{lemma}
Any element of the families $\{\mathbf{Y}_l;\ l \in \cL\}$ and  $\{\mathbf{X}_k;\ k \in \cK\}$  is a polytope.
\end{lemma}
\textbf{Proof:} Since, by definition,
\[Y_\ell=\left\{y \in \cY\ \mbox{s.t.}\  x_\ell \in \arg\max_{\ell' \in \cL} \inf_{U \in \Phi(y)} \langle x_{\ell'}, U\rangle \right\}\] is a polytope of $\mathds{R}^{B}$, there exists a finite family $\left\{b_t \in \mathds{R}^{B}, c_t \in \mathds{R}; t \in \mathcal{T}_\ell\right\}$ such that
 \[Y_\ell=\bigcap_{t \in T_\ell} \Big\{y \in \cY\ \mbox{s.t.}\ \langle y,b_t\rangle \leq c_t\Big\}\ .\]
 Therefore, $\mathbf{Y}_\ell$ is also a polytope of $\mathds{R}^K$ as it can be written as
 \[\mathbf{Y}_\ell=\bigcap_{t \in T_\ell} \Big\{\mathbf{y} \in \Delta(\cK)\ \mbox{s.t.}\ \langle \mathds{E}_{\mathbf{y}}[y],b_t\rangle \leq c_t\Big\}=\bigcap_{t \in T_\ell} \Big\{\mathbf{y} \in \Delta(\cK)\ \mbox{s.t.}\ \left\langle \mathbf{y}, \left(\langle y_k, b_t\rangle\right)_{k \in \cK}\right\rangle \leq c_t \Big\}.\]
Similar arguments hold for   $\{\mathbf{X}_k;\ k \in \cK\}$. $\hfill \Box$

\medskip

Using this important property, we can generalize the LH-algorithm to games with uncertainties satisfying some non-degeneracy assumptions.

\begin{theorem}
If  $\{\mathbf{Y}_\ell;\ \ell \in \cL\}$ and  $\{\mathbf{X}_k;\ k \in \cK\}$ satisfy Assumption~\ref{hypoLH}, then any end-point of Lemke-Howson algorithm induces a Nash equilibrium of $\Gamma$.
\end{theorem}
\textbf{Proof:} If $\{\mathbf{Y}_\ell;\ \ell \in \cL\}$ and  $\{\mathbf{X}_k;\ k \in \cK\}$ satisfy the non-degeneracy Assumption~\ref{hypoLH}, any end point of the LH-algorithm is fully labelled, hence a Nash equilibrium of $\Gamma$. $\hfill \Box$

\begin{remark}
It is not compulsory to use the induced polytopial complexes of $\Delta(\cL)$ and $\Delta(\cK)$. One can work directly in $\cX$ and $\cY$ by considering the projection of the skeleton of the complexes $\{\mathbf{Y}_\ell;\ \ell \in \cL\}$ and $\{\mathbf{X}_k;\ k \in \cK\}$ onto them. However, the graphs generated might not be planar and there are, at first glance, no guarantee that the LH-algorithm will work. In the proof of Theorem~\ref{mainresult}, it is a lifting of the problem that ensures that graphs are planar.
\end{remark}

The fact that there was an odd number of Nash equilibria in the game of Example~\ref{EX:SansSemiStand} (continued in Section~\ref{examplesemistandardnecsuite} below) is therefore not surprising; with full monitoring and the non-degeneracy assumption, this can be proved using the LH-algorithm.  Therefore, as soon as $\{\mathbf{Y}_\ell;\ \ell \in \cL\}$ and  $\{\mathbf{X}_k;\ k \in \cK\}$ satisfy this assumption, there will exist an odd number of fully labelled points in $\Delta(\cL)\times\Delta(\cK)$ inducing Nash equilibria.

\medskip

In some cases,  the main argument of the proof of Theorem~\ref{mainresult} can be rephrased as follows. The game $\Gamma$ is, in fact, equivalent to a game $\widehat{\Gamma}$ with full monitoring, with  action spaces $\cL$ and $\cK$ and with payoffs defined in a arbitrary way so that the polytopial complexes induced by the  best-replies  areas coincide with $\{\mathbf{Y}_\ell;\ \ell \in \cL\}$ and $\{\mathbf{X}_k;\ k \in \cK\}$. However, the existence of such abstracts payoffs might not be ensured in  general  (or it can depend on the duplication of the mixed actions chosen, see Example~\ref{examplefinal}). Anyway, whenever it is possible, it is again almost instantaneous to understand that Nash equilibria of $\Gamma$ and $\widehat{\Gamma}$  coincide.

\begin{example}\label{examplefinal}
Consider the  game defined by, respectively, the following payoffs  and signal (in $\R^2$) matrices for the row player:
\begin{center}
\begin{tabular}{c|c|c|c|c|p{2cm}c|c|c|c|c|}
\multicolumn{1}{c}{} &\multicolumn{1}{c}{$L$} & \multicolumn{1}{c}{$M$} & \multicolumn{1}{c}{$C$} &\multicolumn{1}{c}{$R$} &&\multicolumn{1}{c}{} &\multicolumn{1}{c}{$L$} & \multicolumn{1}{c}{$M$} & \multicolumn{1}{c}{$C$} &\multicolumn{1}{c}{$R$}\\
\cline{2-5}\cline{8-11}
$T$ & 4 & 4& 4& 0& & $T$ & (0,0) & (0,1) & (1,0) & (1,1)\\
\cline{2-5}\cline{8-11}
$B$ & 3 & 3& 3& 3& & $B$ & (0,0) & (0,1) & (1,0) & (1,1)\\
\cline{2-5}\cline{8-11}
\end{tabular}
\end{center}
Given the signal $(\alpha,\beta) \in [0,1]^2$, the best response is $B$ if $\alpha$ and $\beta$ are both bigger than $0.25$ and the best response is $T$ is either $\alpha$ or $\beta$ is smaller than $0.25$. Therefore $Y_B$ is convex but $Y_T$ is not (but it is the union of two polytopes). 

Assume that the column player has a full monitoring and that his four action might be best responses, then $\mathbf{Y}_T$ is not convex and the decomposition $\{\mathbf{Y}_B,\mathbf{Y}_T\}$ cannot be induced by some equivalent game with full monitoring.

\medskip

On the other hand, one can find a  decomposition of $Y_T$ into two polytopes,  namely $Y_{T_1}= H^{-1}\Big(\{(\alpha,\beta)\in[0,1]^2\ \mbox{s.t.}\ \alpha \leq \min( 0.25, \beta)\}\Big)$ and similarly  $Y_{T_2}= H^{-1} \Big( \{ (\alpha,\beta)\in[0,1]^2\ \mbox{s.t.}\ \beta \leq \min(0.25, \alpha) \} \Big)$. It is easy to see that $\{\mathbf{Y}_{T_1},\mathbf{Y}_{T_2},\mathbf{Y}_B\}$ can be induced by some completely auxiliary game with full monitoring -- this decomposition is said to be \textsl{regular}, see~\cite[Definition 5.3 and page 132]{Zie95}. And with respect to this decomposition, $\mathbf{x} \in \Delta(\{T_1,T_2,B\})$ induces the mixed action $x \in \Delta(\{T,B\})$ defined by $x[T]=\mathbf{x}[T_1]+\mathbf{x}[T_2]$.
\end{example}

\section{Examples  with partial monitoring or in robust games}\label{examplesemistandardnecsuite}
Consider again Example~\ref{EX:SansSemiStand}. The polytopial complexes $\{Y_\ell;\ \ell \in \cL\}:=\{Y_B,Y_M,Y_T\}$  and $\{X_k;\ k \in \cK\}:=\{X_C,X_L\}$ are represented in the following figure~\ref{figure}.

\begin{figure}[h!]
\centering
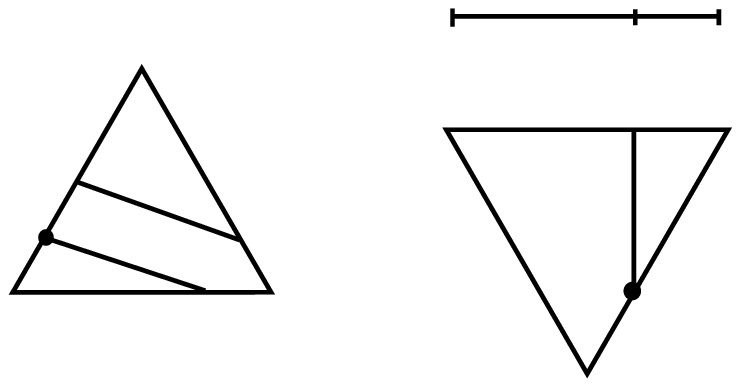
\caption{On the left $\cX$ and on the right $\cY$ and $\Delta(\cK)$ with their complexes.}
\label{figure}
\end{figure}

In order to describe how the LH-algorithm works, we will denote a vertex of the product graph by the cartesian product of its labels (in this example  the set of labels is $\{T,B,M,L,R,C\}$); for example the vertex represented with a black dot in figure~\ref{figure} is denoted by $\{R,T,M\}\times\{B,C,L\}$.

The first step in the LH-algorithm is to drop one label arbitrarily;  If the label $M$ is dropped then the first vertex visited by the algorithm is $ \{L,R,C\}\times\{B,T,C\}$. The label $C$ appears twice, so in order to get rid of one of them, the algorithm chooses at the next step  the vertex $ \{L,R,B\}\times\{B,T,C\}$ and the following vertex is $ \{L,R,B\}\times\{M,T,C\}$. It is fully labelled, thus an end point of the algorithm, hence $(B,C) \in \cX\times\cY$ is a pure Nash equilibrium of $\Gamma$.

Similarly, If $T$ is dropped at the first stage, then the first vertex is $ \{L,R,C\}\times\{B,M,L\}$ and the second $ \{R,C,T\}\times\{B,M,L\}$. So $(T,L)$ is also a pure Nash equilibrium of $\Gamma$.

Starting  again from this point and dropping the label $C$ makes the LH-algorithm visit $ \{R,T,M\}\times\{B,M,L\}$, and then $ \{R,T,M\}\times\{B,L,C\}$ which is also a Nash equilibrium. It corresponds to $(\mathbf{x},\mathbf{y})=(1/3T+2/3M,3/4L+1/4C) \in \Delta(\cL)\times \Delta(\cK)$ which induces $(x,y)=(2/3T+1/3B,3/4L+1/4C)$ which is a (mixed) Nash equilibrium of $\Gamma$.

One can check the remaining vertices of the product graph to be convinced that there does not exist any more equilibria.

\bigskip

We now quickly treat the case of \textsl{robust games} where players observe their opponents actions but their payoff mapping is unknown; the only information is that $u$ belongs to some polytope $\mathbf{U}$ (and $v$ to some $\mathbf{V}$). Then under those assumptions uncertainties correspondence $\Phi$ and $\Psi$ might not be piece-wise linear.

\begin{example}
Assume that the payoff matrix of player 1 belongs to the convex hull of the following two matrices, i.e., $\mathbf{U}=\{\lambda u_1+(1-\lambda)u_2 ; \lambda \in [0,1]\}$ with
\begin{center}
\begin{tabular}{p{0.2cm}cp{0.2cm}p{1.5cm}p{0.2cm}c}
$u_1=$&
\begin{tabular}{c|c|c|}
\multicolumn{1}{c}{}&\multicolumn{1}{c}{$L$}&\multicolumn{1}{c}{$R$}\\
\cline{2-3}
$T_1$ & 1 & 0\\
\cline{2-3}
$T_2$ & 0 & 0\\
\cline{2-3}
$B_1$ & 0 & 1\\
\cline{2-3}
$B_2$ & 0 & 0\\
\cline{2-3}
\end{tabular}
&&and&$u_2=$&
\begin{tabular}{c|c|c|}
\multicolumn{1}{c}{}&\multicolumn{1}{c}{$L$}&\multicolumn{1}{c}{$R$}\\
\cline{2-3}
$T_1$ & 0 & 0\\
\cline{2-3}
$T_2$ & 1 & 0\\
\cline{2-3}
$B_1$ & 0 & 0\\
\cline{2-3}
$B_2$ & 0 & 1\\
\cline{2-3}
\end{tabular}
\end{tabular}
\end{center}
Then for any $y \in [0,1]$, 
\[\Phi\Big(yL+(1-y)R\Big)=\left\{ \left(\begin{array}{c}
 y\lambda \\
 y(1-\lambda) \\
 (1-y)\lambda\\
 (1-y)(1-\lambda)\\
\end{array}\right)\ ; \ \lambda \in [0,1]\right\}\]
which is not piece-wise linear in $y$. Indeed $\Phi(yL+(1-y)R)$ can be seen as the set of product probability distributions over $\{T,B\}\times\{1,2\}$ with first marginal $yT+(1-y)B$.
\end{example}

As a consequence, Lemma~\ref{lemmalineargeneral} might not hold so Lemke-Howson algorithm can not, in general, be extended (see e.g.,~\cite[Section 5]{AghBer06} for alternative technics). On the other hand, if both players have only 2 actions, then it is not difficult to see that $\Phi$ and $\Psi$ are piecewise linear (as they cannot \textsl{turn} as in higher dimensions); so in that specific case, our results extend.

\section*{Some open questions}
Important questions remains open. We have shown that under some regularity (or non-degeneracy) assumption on the decomposition into best reply areas, Nash equilibria are induced by an odd number of points. The characterization of such games (maybe as a large semi-algebraic class or such that a game chosen uniformly in some open ball satisfy it with probability one) appears to be a real challenging problem here. With full monitoring, one just has to check that vectors $u(\cdot,b)$ and $v(a,\cdot)$ are in some generic position. With partial monitoring, one must first control the fact that $x_\ell$ and $y_k$ are themselves in generic position, then that $\mathbf{Y}_\ell$ and $\mathbf{X}_k$ also satisfy  regularity conditions; moreover, genericity can  be described with respect to the mappings $u$ and $v$ (as in full monitoring) or to $H$ and $M$, or  to simultaneously all of them. Answering this question will most probably require a deeper understanding of how normal cones evolve with $u,v, H$ and $M$.

Other questions concern wether index and stability of these equilibria can be defined and studied, see~\cite{von05}:  for instance, we can  wonder which equilibria remains in a neighborhood of a given game. The complexity of computing these equilibria, and wether it is in the same class than with full monitoring~\cite{Pap07},  must also be addressed.

\bigskip

\textbf{Acknowledgments:} I am grateful to S.\ Sorin for his -- as always -- useful comments and to F.\ Riedel, B.\ von Stengel and G.\ Vigeral  for their wise remarks.

\end{document}